\documentclass[prb,reprint,notitlepage,amsmath,amsfonts,amssymb,floatfix]{revtex4-1}
\usepackage{graphicx}
\usepackage{bm}

\begin{document}
\title{Projected Hartree-Fock as a Polynomial of Particle-Hole Excitations and Its Combination With Variational Coupled Cluster Theory}
\author{Yiheng Qiu}
\affiliation{Department of Chemistry, Rice University, Houston, TX, USA 77005-1892}

\author{Thomas M. Henderson}
\affiliation{Department of Chemistry, Rice University, Houston, TX, USA 77005-1892}
\affiliation{Department of Physics and Astronomy, Rice University, Houston, TX, USA 77005-1892}

\author{Gustavo E. Scuseria}
\affiliation{Department of Chemistry, Rice University, Houston, TX, USA 77005-1892}
\affiliation{Department of Physics and Astronomy, Rice University, Houston, TX, USA 77005-1892}
\date{\today}

\begin{abstract}
Projected Hartree-Fock theory provides an accurate description of many kinds of strong correlation but does not properly describe weakly correlated systems.  Coupled cluster theory, in contrast, does the opposite.  It therefore seems natural to combine the two so as to describe both strong and weak correlations with high accuracy in a relatively black-box manner.  Combining the two approaches, however, is made more difficult by the fact that the two techniques are formulated very differently.  In earlier work, we showed how to write spin-projected Hartree-Fock in a coupled-cluster-like language.  Here, we fill in the gaps in that earlier work.  Further, we combine projected Hartree-Fock and coupled cluster theory in a variational formulation and show how the combination performs for the description of the Hubbard Hamiltonian and for several small molecular systems.
\end{abstract}
\maketitle

\section{Introduction}
The successes of traditional single-reference coupled cluster theory\cite{Coester1958,Cizek1966,Paldus1999,Bartlett2007,ShavittBartlett} are manifold, and the method is rightly held as the gold standard of quantum chemistry.  These successes, however, come with an important caveat: affordable coupled cluster calculations are generally only accurate for weakly correlated systems.  For strongly correlated problems involving more than two strongly correlated electrons, coupled cluster theory is significantly less accurate and often fails badly.  

Several techniques exist to address the strongly correlated regime.  Among the simplest computationally is the use of symmetry projected mean-field methods to which we will refer generically as projected Hartree-Fock (PHF).\cite{Lowdin55c,Ring80,Blaizot85,Schmid2004,PHF}   The idea is also simple: when electrons become strongly correlated, mean-field methods such as Hartree-Fock theory no longer provide even qualitatively reasonable descriptions of their behavior (which is, at heart, why coupled cluster theory fails).  However, mean-field methods typically signal their own breakdown by artificially breaking some or even all symmetries of the problem.  These symmetry-broken mean-field methods often contain valuable information.  For example, unrestricted Hartree-Fock (UHF) correctly localizes the two electrons to separate nuclei in the dissociation of H$_2$ at the cost of good spin (and spatial) symmetry, while restricted Hartree-Fock (RHF) imposes spin symmetry and in doing so delivers unphysical results.  Of course the broken symmetries of UHF are not broken in the exact wave function, but projecting out the component of the UHF wave function in spin-projected UHF (SUHF) restores those symmetries but retains the advantages of UHF and dissociates H$_2$ correctly.  While SUHF in the form introduced by L\"owdin as extended Hartree-Fock\cite{Lowdin55c} is computationally cumbersome, it can be reformulated in a way which has mean-field computational scaling using integration over generalized symmetry coherent states.\cite{Schmid2004,PHF}

Much of our work in the past year has revolved around combining these two approaches.\cite{Degroote2016,Qiu2016,WahlenStrothman2016,Hermes2017}  The main challenge is that the two techniques use fundamentally very different frameworks.  Coupled cluster theory introduces a similarity-transformed Hamiltonian constructed via particle-hole excitation operators, and solves this similarity-transformed Hamiltonian in a subspace of the full Hilbert space.  In contrast, PHF is a variational approach which minimizes the expectation value of a projected mean-field state with respect to the identity of the broken symmetry determinant from which the PHF wave function is built.

Recently, we showed an analytic resummation of the SUHF wave function in terms of particle-hole excitations constructed out of a symmetry adapted RHF mean-field determinant.\cite{Qiu2016}  While this is an important piece of the puzzle, we did not show any results for the combination of PHF and coupled cluster.  This manuscript remedies that deficiency and fills in details missing from our earlier work.

\section{The SUHF Polynomial}
Let us begin by proving the main contention of Ref. \onlinecite{Qiu2016}, that SUHF is a polynomial of double excitations when written in the RHF basis.  While this result is novel, it was inspired by results obtained in Ref. \onlinecite{Piecuch1996}; see also Ref. \onlinecite{Stuber2016}.

Conventionally, we would write the spin-projected UHF wave function in terms of an integration of rotation operators acting on the UHF determinant:
\begin{equation}
|\mathrm{SUHF}\rangle = \frac{1}{8 \, \pi^2} \, 
   \int\limits_0^{2 \, \pi} \mathrm{d}\alpha \, 
   \int\limits_0^\pi \, \sin(\beta) \, \mathrm{d}\beta \, 
   \int\limits_0^{2 \, \pi} \mathrm{d}\gamma \, R |\mathrm{UHF}\rangle,
\end{equation}
where we have already specialized to the case of projecting a singlet spin, and have defined the spin rotation operator 
\begin{equation}
R = \mathrm{e}^{\mathrm{i} \, \alpha \, S_z} \, 
    \mathrm{e}^{\mathrm{i} \, \beta \, S_y} \, 
    \mathrm{e}^{\mathrm{i} \, \gamma \, S_z}.
\label{Eqn:DefR}
\end{equation}
It is helpful to note that the integrations with respect to $\alpha$ and $\gamma$ amount to projectors onto $S_z$ eigenstates with eigenvalue 0, and that the UHF determinant is an eigenfunction of this projector, so we can write more simply
\begin{equation}
|\mathrm{SUHF}\rangle = \frac{1}{2} \, P_{S_z = 0} \, \int\limits_0^\pi \, \sin(\beta) \, \mathrm{d}\beta \, \mathrm{e}^{\mathrm{i} \, \beta \, S_y} \, |\mathrm{UHF}\rangle.
\end{equation}

For our purposes, it will be convenient to write the UHF determinant as a Thouless transformation\cite{Thouless1960} from some RHF determinant:
\begin{subequations}
\begin{align}
|\mathrm{UHF}\rangle &= \mathrm{e}^{T_1 + U_1} \, |\mathrm{RHF}\rangle,
\\
T_1 &= \sum_{ia} t_i^a \, \left(c_{a_\uparrow}^\dagger \, c_{i_\uparrow} + c_{a_\downarrow}^\dagger \, c_{i_\downarrow}\right),
\\
U_1 &= \sum_{ia} u_i^a \, \left(c_{a_\uparrow}^\dagger \, c_{i_\uparrow} - c_{a_\downarrow}^\dagger \, c_{i_\downarrow}\right).
\end{align}
\end{subequations}
Here and throughout, spatial orbitals $i$, $j$, $k$, \ldots are occupied and $a$, $b$, $c$, \ldots are virtual.  Because the operator $T_1$ preserves spin, it commutes with the spin projection operator and we will disregard it in what follows so as to simplify the derivation.  Note that a general Thouless transformation has two more pieces which mix $\uparrow$ and $\downarrow$ spin, but as these pieces yield generalized Hartree-Fock rather than UHF, we omit them here.

With the UHF determinant represented as a Thouless transformation from the RHF determinant, we can write the SUHF state as
\begin{equation}
|\mathrm{SUHF}\rangle =\frac{1}{2} \, P_{S_z = 0} \, \int\limits_0^\pi \, \sin(\beta) \, \mathrm{d}\beta \, \mathrm{e}^{\tilde{U}_1} \, |\mathrm{RHF}\rangle,
\end{equation}
where we have introduced
\begin{equation}
\tilde{U}_1 = \mathrm{e}^{\mathrm{i} \, \beta \, S_y} \, U_1 \, \mathrm{e}^{-\mathrm{i} \, \beta \, S_y} 
\label{Eqn:UTilde}
\end{equation}
and have used the fact that the RHF determinant is an eigenfunction of $S_y$ with eigenvalue 0.  Using the representation of $S_y$ in terms of fermionic creation and annihilation operators and taking advantage of the Baker-Campbell-Hausdorff commutator expansion, we find that
\begin{subequations}
\begin{align}
\tilde{U}_1
 &= \sum_{ia} u_i^a \, \Big[ \cos(\beta) \, \left(c_{a_\uparrow}^\dagger \, c_{i_\uparrow} - c_{a_\downarrow}^\dagger \, c_{i_\downarrow}\right) 
\\
 &\hspace{15mm}- \sin(\beta) \, \left(c_{a_\uparrow}^\dagger \, c_{i_\downarrow} + c_{a_\downarrow}^\dagger \, c_{i_\uparrow}\right)\Big]
\nonumber
\\
 &= \cos(\beta) \, U_1 - \sin(\beta) \, \left(U_1^{(+)} + U_1^{(-)}\right)
\end{align}
\end{subequations}
where we have introduced
\begin{subequations}
\begin{align}
U_1^{(+)} &= \sum_{ia} u_i^a \, c_{a_\uparrow}^\dagger \, c_{i_\downarrow},
\\
U_1^{(-)} &= \sum_{ia} u_i^a \, c_{a_\downarrow}^\dagger \, c_{i_\uparrow}.
\end{align}
\end{subequations}
Note that $U_1^{(+)}$ increases the $S_z$ eigenvalue of an $S_z$ eigenstate by 1, while $U_1^{(-)}$ decreases it.

With these ingredients in hand, the SUHF wave function is
\begin{align}
|\mathrm{SUHF}\rangle
 &= P_{S_z = 0} \, \sum_{n=0}^{\infty} \sum_{k=0}^n \frac{1}{n!} \, {n \choose k} I_{n,k} 
\\
 &\hspace{1cm}\times U_1^{n-k} \, \left(U_1^{(+)} + U_1^{(-)}\right)^k |\mathrm{RHF}\rangle
\nonumber
\end{align}
where $I_{n,k}$ is the integral
\begin{equation}
I_{n,k} = \frac{1}{2} \, \left(-1\right)^k \, \int\limits_{-1}^1 \mathrm{d}\cos(\beta) \, \cos(\beta)^{n-k} \, \sin(\beta)^k.
\end{equation}
The $S_z$ projector means that in $(U_1^{(+)} + U_1^{(-)})^k$ we must have as many powers of $U_1^{(+)}$ as of $U_1^{(-)}$, which means $k$ must be even, so we write $k = 2 j$.  The integral $I_{n,2j}$ vanishes unless $n$ is also even, and we write $n = 2m$.  Using these facts simplifies the wave function to
\begin{align}
|\mathrm{SUHF}\rangle
 &= \sum_{m=0}^{\infty} \sum_{j=0}^m \frac{1}{(2m)!} \, {2m \choose 2j} \, {2j \choose j} \, I_{2m,2j} 
\\
 &\hspace{1cm}\times U_1^{2m-2j} \, \left(U_1^{(+)} \, U_1^{(-)}\right)^j |\mathrm{RHF}\rangle
\nonumber
\end{align}

We can analytically evaluate $I_{2m,2j}$ as an Euler Beta function:
\begin{equation}
\frac{1}{2} \, \int\limits_{-1}^1 \, \mathrm{d}x \, x^{2m-2j} \, (1-x^2)^j = \frac{j! \, \Gamma(m-j+1/2)}{2 \, \Gamma(m+3/2)},
\end{equation}
where we have written $x = \cos(\beta)$, so that
\begin{align}
|\mathrm{SUHF}\rangle
 &= \sum_{m=0}^\infty \sum_{j=0}^m \frac{1}{(2m)!} \, {2m \choose 2j} \, {2j \choose j} \,  \frac{j! \, \Gamma(m-j+1/2)}{2 \, \Gamma(m+3/2)}
\nonumber
\\
 &\hspace{5mm}\times U_1^{2m-2j} \, \left(U_1^{(+)} \, U_1^{(-)}\right)^j |\mathrm{RHF}\rangle
\end{align}

This result is deceptively formidable.  The combinatorial factors, however, simplify enormously.  Using the fact that
\begin{equation}
\Gamma(m+1/2) = \sqrt{\pi} \, \frac{(2m)!}{m! \, 4^m}
\end{equation}
allows us to write
\begin{align}
 \frac{1}{(2m)!} \, {2m \choose 2j} \, & {2j \choose j} \,  \frac{j! \, \Gamma(m-j+1/2)}{2 \, \Gamma(m+3/2)}
\\
 &= \frac{1}{(2m+1)!} \, {m \choose j} \, 4^j
\nonumber
\end{align}
from which we obtain simply
\begin{subequations}
\begin{align}
|\mathrm{SUHF}\rangle 
 &= \sum_{m=0}^{\infty} \frac{1}{(2m+1)!} \, \sum_{j=0}^m {m \choose j} 
\\
 &\times \left(U_1^2\right)^{m-j} \, \left(4 \, U_1^{(+)} \, U_1^{(-)}\right)^j |\mathrm{RHF}\rangle
\nonumber
\\
 &= \sum_{m=0}^\infty \frac{1}{(2m+1)!} \, \left(6 \, K_2\right)^m |\mathrm{RHF}\rangle
\label{Eqn:DefF}
\\
 &= \frac{\sinh(\sqrt{6 \, K_2})}{\sqrt{6 \, K_2}} |\mathrm{RHF}\rangle
\end{align}
\end{subequations}
where we have defined
\begin{equation}
6 \, K_2 = U_1^2 + 4 \, U_1^{(+)} \, U_1^{(-)}
\end{equation}
so that the polynomial $F(K_2) = \sum 1/(2m+1)! \, (6 \, K_2)^m$ implicitly defined in Eqn. \ref{Eqn:DefF} begins with $1 + K_2 + \ldots$.

Finally, it is useful to write $K_2$ in terms of the amplitudes in $U_1$ and excitation operators.  In agreement with Ref. \onlinecite{Piecuch1996}, we find that
\begin{subequations}
\begin{align}
K_2 &= \frac{1}{6} \, \left(U_1^2 + 4 \, U_1^{(+)} \, U_1^{(-)}\right)
\\
    &= -\frac{1}{6} \, \sum_{ijab} \left(u_i^a \, u_j^b + 2 \, u_i^b \, u_j^a\right) E_a^i \, E_b^j
\end{align}
\end{subequations}
where $E_a^i$ and $E_b^j$ are unitary group generators
\begin{equation}
E_a^i = c_{a_\uparrow}^\dagger \, c_{i_\uparrow} + c_{a_\downarrow}^\dagger \, c_{i_\downarrow}.
\end{equation}

Taken all together, these results demonstrate that the SUHF wave function can be written as a polynomial of double excitations out of the RHF ground state where the double excitation operators are spin adapted and factorizable.  One should not forget that in addition, spin-adapted single excitations must be included.

\section{Symmetry-Projected Extended Coupled Cluster}
\begin{figure}[t]
\includegraphics[width=\columnwidth]{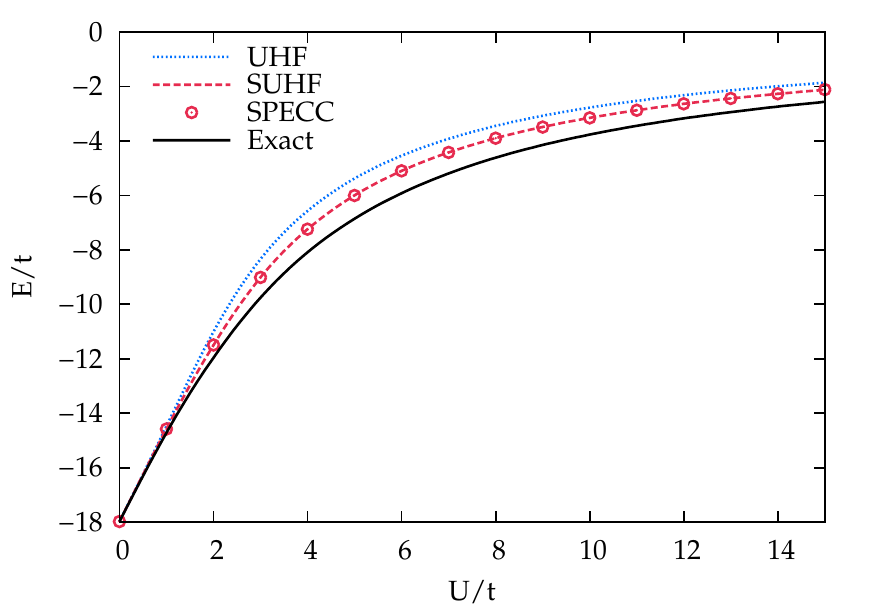}
\caption{Total energies in the 14-site 1/2-filled Hubbard Hamiltonian with periodic boundary conditions.  UHF and SUHF are variationally optimized, while SPECC stands for the method described in and around Eqn. \ref{Eqn:ESPECC}.
\label{Fig:SPECC}}
\end{figure}

Once we have the SUHF wave function in this particle-hole language, we might naturally wish to optimize the parameters defining it and extract the ground state energy.  To do so, Ref. \onlinecite{Qiu2016} defined an energy expression
\begin{equation}
E = \langle \mathrm{RHF}| G(L_2) \, F^{-1}(K_2)  \, H \, F(K_2) |\mathrm{RHF}\rangle
\label{Eqn:ESPECC}
\end{equation}
where $L_2$ is an operator of the same form as $K_2^\dagger$ but expressed with a different set of amplitudes $v^i_a$, and where $G(L_2)$ is a polynomial which seeks to make the left-hand state look as much like the adjoint of the right-hand-state as possible.  The resulting energy is made stationary with respect to the amplitudes $u_i^a$ and $v^i_a$; additionally, we may include spin-adapted single excitations in the traditional coupled-cluster style.  A characteristic result is shown in Fig. \ref{Fig:SPECC} where we demonstrate that with suitably optimized $G(L_2)$ we closely reproduce the traditional PHF energy.  This good agreement between SUHF and our similarity-transformed approach demonstrates the correctness of our energy expression and amplitude equations.

Evaluating the energy above is not entirely trivial.  We will sketch our technique here and provide full details in the supplementary material.

Our energy expression is
\begin{equation}
E = \sum_{lmn} a_l \, \bar{c}_m \, c_n \, \langle \mathrm{RHF}| L_2^l \, K_2^m  \, H \, K_2^n |\mathrm{RHF}\rangle
\end{equation}
where $a_l$, $\bar{c}_m$, and $c_n$ are the coefficients in the polynomials $G(L_2)$, $F^{-1}(K_2)$, and $F(K_2)$, respectively.  The key task is thus to evaluate matrix elements
\begin{equation}
E_{lmn} = \langle \mathrm{RHF}| L_2^l \, K_2^m  \, H \, K_2^n |\mathrm{RHF}\rangle.
\end{equation}
In order to evaluate this matrix element efficiently, we replace $L_2$ and $K_2$ with
\begin{subequations}
\begin{align}
P_0 \, K_2 \, P_0 &= \frac{1}{2} \, P_0 \, U_1^2 \, P_0,
\\
P_0 \, L_2 \, P_0 &= \frac{1}{2} \, P_0 \, V_1^2 \, P_0,
\end{align}
\end{subequations}
where $P_0$ is the projection operator onto $S=0$ and $V_1$ is explicitly
\begin{equation}
V_1 = \sum_{ia} v^i_a \, \left(c_{i_\uparrow}^\dagger \, c_{a_\uparrow} - c_{i_\downarrow}^\dagger \, c_{a_\downarrow}\right).
\end{equation}
We can introduce as many projection operators $P_0$ as we need because they are idempotent, commute with $K_2$, $L_2$, and $H$, and 
\begin{equation}
P_0 |\mathrm{RHF}\rangle = |\mathrm{RHF}\rangle.
\end{equation}
Most of these projection operators can then be eliminated by using the fact (demonstrated in the supplementary material) that
\begin{equation}
P_0 \, U_1^a \, P_0 \, U_1^b \, P_S = \epsilon_{a,b}^S \, P_0 \, U_1^{a+b} \, P_S
\label{Eqn:EpsReln}
\end{equation}
where $P_S$ is the projector onto spin $S$ with $S_z = 0$.  A consequence of this relation is that
\begin{equation}
P_0 \, K_2^n \, P_0 = p_n \, P_0 \, U_1^{2n} \, P_0
\end{equation}
where $p_n$ is a simple coefficient; in fact, 
\begin{equation}
p_n = \frac{2n+1}{6^n}
\end{equation}
as can be readily derived from the fact that
\begin{align}
|\mathrm{SUHF}\rangle 
  &= P_0 \mathrm{e}^{U_1} \, P_0 |\mathrm{RHF}\rangle
\\
  &= P_0 \, F(K_2) \, P_0 |\mathrm{RHF}\rangle.
\nonumber
\end{align}
Similar results of course hold for $V_1$ and $L_2$.

Using these relations gives us
\begin{widetext}
\begin{subequations}
\begin{align}
E_{lmn}
 &= p_l  \, p_m  \, p_n \, \langle \mathrm{RHF} |V_1^{2l} \, P_0 \, U_1^{2m} \, P_0 \, H \, U_1^{2n} |\mathrm{RHF}\rangle
\\
 &= p_l  \, p_m  \, p_n \, \sum_{k=0}^4 \, {2n \choose k} \langle \mathrm{RHF} |V_1^{2l} \, P_0 \, U_1^{2m} \, P_0 \, U_1^{2n-k} \, \left(H \, U_1^k\right)_c |\mathrm{RHF}\rangle
\\
 &= p_l  \, p_m  \, p_n \, \sum_{k=0}^4 \, \sum_S {2n \choose k} \langle \mathrm{RHF} |V_1^{2l} \, P_0 \, U_1^{2m} \, P_0 \, U_1^{2n-k} \, P_S \, \left(H \, U_1^k\right)_c |\mathrm{RHF}\rangle
\\
 &=  p_l \, p_m \, p_n \, \sum_{k=0}^4 \sum_S {2n \choose k} \, \epsilon^S_{2m,2n-k} \,  \langle \mathrm{RHF} | V_1^{2l} \, P_0 \, U_1^{2m+2n-k} \, P_S \, \left(H \, U_1^k\right)_c|\mathrm{RHF}\rangle.
\end{align}
\end{subequations}
\end{widetext}
Here, the subscript ``c'' stands for the connected component.  In going from the first line to the second we have used
\begin{equation}
H \, U_1^{2n} = \sum_{k=0}^4 \, {2n \choose k} U_1^{2n-k} \, \left(H \, U_1^k\right)_c.
\end{equation}
In going from the second to the third we have introduced the resolution of unity in the form $1 = \sum_S P_S$, and in going from the third line to the fourth we have used Eqn. \ref{Eqn:EpsReln}.  In principle $S$ must run over all possible spin states that can be constructed from $\left(H \, U_1^k\right)_c |\mathrm{RHF}\rangle$; in practice, we need only $S$ = 0, 1, or 2.

At this point, we can replace the projection operators by integrations over a grid, following our standard techniques:
\begin{widetext}
\begin{align}
E_{lmn}
= p_l \, p_m \, p_n \, \sum_{k=0}^4 \sum_S {2n \choose k} \, & \epsilon^S_{2m,2n-k} \,  \left(2 \, S + 1\right) \int \frac{\mathrm{d}\Omega_1}{8 \, \pi^2} \frac{\mathrm{d}\Omega_2}{8 \, \pi^2} \, D^S_{0,0}(\Omega_2)
\\
 &\times \langle \mathrm{RHF} | \tilde{V}_1(\Omega_1)^{2l} \, U_1^{2m+2n-k} \, \left[H \, \tilde{U}_1(\Omega_2)^k\right]_c |\mathrm{RHF}\rangle
\nonumber
\end{align}
\end{widetext}
where $\Omega$ is a compact notation for the Euler angles over which we integrate and $\tilde{U}_1$ and $\tilde{V}_1$ are the rotated $U_1$ and $V_1$ operators (\textit{c.f.} Eqn. \ref{Eqn:UTilde}, but note that we must now also include rotations with angles $\alpha$ and $\gamma$ in the rotation operator of Eqn. \ref{Eqn:DefR}).

Though our expression is somewhat cumbersome, it can now be evaluated.  The idea is to break it down into a sum of products of connected components; there are $\mathcal{O}(N^2)$ such connected components needed to evaluate the energy for all $l$, $m$, and $n$, each of which can be evaluated in $\mathcal{O}(N^4)$ or less.  We thus evaluate and store all possible connected components, then put them together to evaluate a given combination of indices $l$, $m$, $n$, $k$, and $S$.

To make this concrete, let us show a single example in which we will suppress the explicit dependence on rotation angles.  One possible term is
\begin{align}
\langle V_1^4 \, U_1^2 \, &\left(H \, U_1^2\right)_c\rangle
\\
 &= 12 \, \langle \left(V_1 \, U_1\right)_c\rangle^2 \, \langle V_1^2 \, \left(H \, U_1^2\right)_c\rangle
\nonumber
\\
 &+ 6 \, \langle \left(V_1^2 \, U_1^2\right)_c\rangle \, \langle V_1^2 \, \left(H \, U_1^2\right)_c\rangle
\nonumber
\\
 &+ 24 \, \langle \left(V_1 \, U_1\right)_c\rangle \, \langle \left(V_1^2 \, U_1\right)_c \, V_1 \, \left(H \, U_1^2\right)_c\rangle
\nonumber
\\
 &+ 6 \, \langle \left(V_1^2 \, U_1\right)^2_c \, \left(H \, U_1^2\right)_c\rangle
\nonumber
\\
 &+ 4 \, \langle \left(V_1^3 \, U_1^2\right)_c \, V_1 \, \left(H \, U_1^2\right)_c\rangle
\nonumber
\end{align}
where each of these pieces can readily be evaluated using diagrammatic techniques.

Note finally that each diagram resolves into the product of a spatial orbital part and a spin part.  The spin integration ultimately can be carried out analytically, as we have done for the SUHF wave function earlier.

In addition to evaluating the energy, we must solve for the amplitudes in $U_1$ and $V_1$.  For the moment, we use a simple steepest-descents-style algorithm and calculate the derivatives of the energy with respect to $U_1$ and $V_1$ amplitudes numerically.

\begin{figure*}[t]
\includegraphics[width=0.48\textwidth]{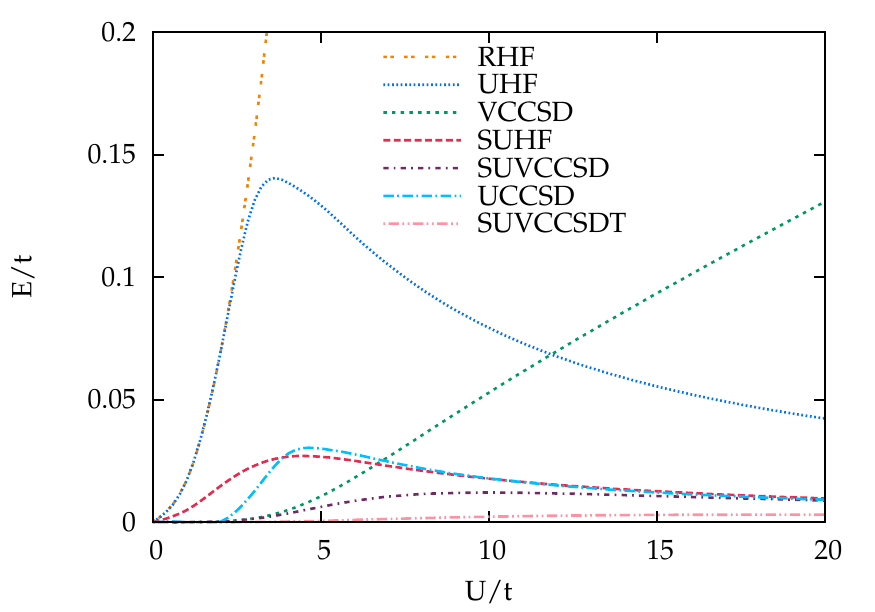}
\hfill
\includegraphics[width=0.48\textwidth]{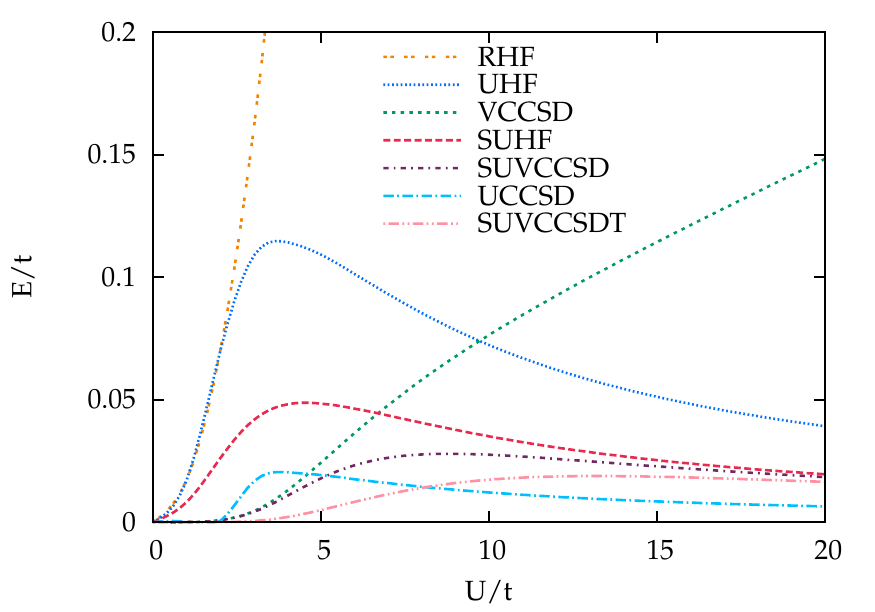}
\caption{Errors per electron with respect to the exact result in half-filled one-dimensional Hubbard rings.  Left panel: 6-site Hubbard model.  Right panel: 10-site Hubbard model.
\label{Fig:Hubbard}}
\end{figure*}

\section{Combining PHF and Coupled Cluster}
Thus far, we have limited ourselves to the PHF ansatz.  While this is useful for the description of static correlation, it is far from ideal for the description of weakly correlated systems, for which we would prefer a combination with coupled cluster theory.  Note that previous work has built on ways to combine PHF with perturbation theory\cite{VanVoorhis2014} or with configuration interaction,\cite{Tenno2016a,Tenno2016b} albeit in a rather different way.

Unfortunately, it is not entirely clear how this is best to be carried out.  The basic idea is straightforward enough: we could construct a similarity-transformed Hamiltonian
\begin{equation}
\bar{H} = \mathrm{e}^{-T} \, F^{-1}(K_2) \, H \, F(K_2) \, \mathrm{e^T}
\end{equation}
and introduce a biorthogonal expression for the energy which we make stationary with respect to the various parameters of the problem.  Thus, for example, we might write
\begin{equation}
E = \langle \mathrm{RHF}| \left(1 + Z\right) \, G(L_2) \, \bar{H} |\mathrm{RHF}\rangle,
\end{equation}
where $Z$ is a sum of de-excitation operators.  Unfortunately, the techniques outlined above cannot avail us once we include genuinely connected double excitations, and evaluating the energy and amplitude equations for this similarity-transform-based combination of PHF and coupled cluster is far from simple.  For the moment, we will avoid these issues by moving to a strictly variational treatment\cite{Noga1988,VanVoorhis2000,Knowles2010,Evangelista2011} as proof of principle.  If the variational method works well, then we have reason to invest the time and effort in a similarity-transformed coupled-cluster-like approach, whereas if the variational treatment fails, then the similarity-transformed approach is unlikely to be successful.

Our variational approach writes
\begin{equation}
|\Psi\rangle = F(K_2) \, \mathrm{e}^T |\mathrm{RHF}\rangle = P_{S = 0} \, \mathrm{e}^{T + U_1} \, |\mathrm{RHF}\rangle,
\end{equation}
defines the energy by the usual expectation value
\begin{equation}
E = \frac{\langle \Psi|H|\Psi\rangle}{\langle \Psi|\Psi\rangle},
\end{equation}
and obtains the amplitudes by making the energy stationary.  We have adopted a full configuration interaction code to do the calculations and have used the conjugate gradients method to solve for the wave function parameters.  Though relatively straightforward to implement, this does limit our results in this manuscript to systems which are sufficiently small that exact results are readily available.  We limit the cluster operator $T$ to single and double substitutions so that we have variational coupled cluster singles and doubles (VCCSD) and, when combined with SUHF, spin projected unrestricted variational coupled cluster singles and doubles SUVCCSD).  We will also report the effects of triple excitations in combination with spin projection to obtain spin projected unrestricted variational coupled cluster singles, doubles, and triples (SUVCCSDT).  This implementation of SUHF agrees with our more conventional integration over symmetry coherent states to the precision with which we solve the respective equations (about ten decimal places).

We must make one cautionary note.  While we can initialize $T = 0$ in SUVCC, we must provide a non-zero initial guess for $U_1$ because $U_1 = 0$ is a solution of the SUHF and SUVCCSD equations.  The results of our calculations depend on the choice of initial guess.  Usually it suffices to initialize $U_1$ to the eigenvector of the SUHF Hessian corresponding to the lowest eigenvalue at $U_1 = 0$.  Occasionally, however, this procedure is insufficient.  One can frequently circumvent this problem by carrying out a UHF calculation first and finding the Thouless transformation between the RHF and UHF solution, though of course this is only helpful if the UHF breaks spin symmetry.  One can also use the Thouless transformation between the RHF determinant and the deformed SUHF determinant, which always exists and which can be found by other means.

\section{Results}
We will first discuss results for the Hubbard model Hamiltonian\cite{Hubbard1963} before turning to selected results for small molecules.  Most calculations were done using in-house code, but the UHF and UHF-based coupled cluster instead used the \texttt{Gaussian} program package.\cite{gdv}

\begin{figure}
\includegraphics[width=\columnwidth]{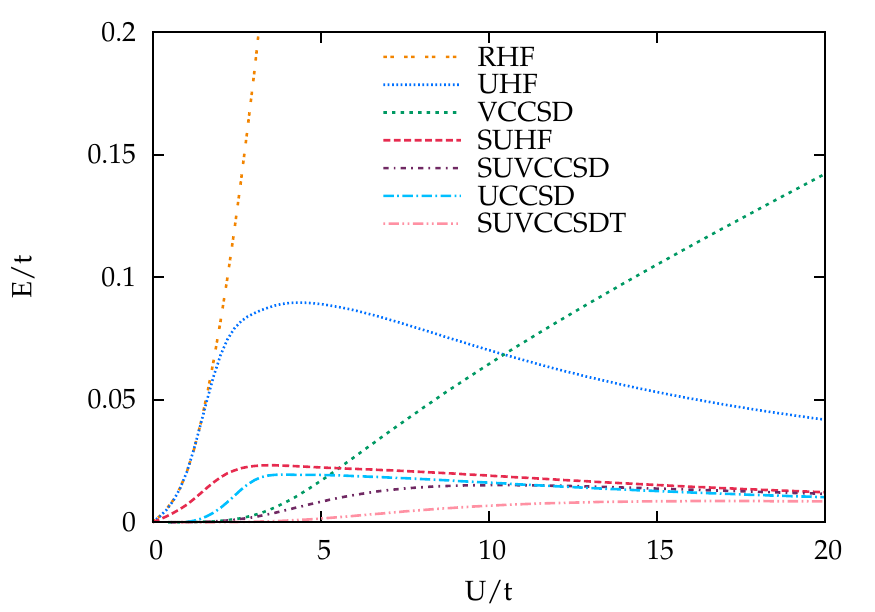}
\caption{Errors per electron with respect to the exact result in the half-filled $2 \times 4$ Hubbard lattice with open boundary conditions.
\label{Fig:Hubbard2x4Half}}
\end{figure}

\subsection{The Hubbard Hamiltonian}
The Hubbard Hamiltonian describes electrons on a lattice.  Electrons can hop from one side to neighboring sites, while two electrons on the same site repel one another.  Mathematically, we write it as
\begin{equation}
H = -t \sum_{\langle ij \rangle} \left(c_{i_\uparrow}^\dagger \, c_{j_\uparrow} + c_{i_\downarrow}^\dagger \, c_{j_\downarrow}\right) + U \sum_i c_{i_\uparrow}^\dagger \, c_{i_\downarrow}^\dagger \, c_{i_\downarrow} \, c_{i_\uparrow}
\end{equation}
where $i$ and $j$ index lattice sites and the notation $\langle ij\rangle$ means the sum runs over sites between which hopping is allowed which, for our purposes, will be nearest-neighbor sites only.  The lattice may be periodic or have open boundaries, and may be one-dimensional or multi-dimensional.  We will limit ourselves to periodic lattices which are generally of more interest; results for open lattices are broadly similar.  Note that the periodicity of the lattice only affects the hopping interaction, so for large $U/t$ there is little practical difference between open and periodic boundary conditions.  Exact results are readily available for one-dimensional periodic lattices thanks to a form of Bethe ansatz,\cite{LiebWu} and high-quality benchmark data is available also for the more physically relevant two-dimensional lattices.\cite{Simons}

For small $U/t$, the system is weakly correlated and can be accurately modeled by traditional coupled cluster on the RHF reference.  As $U/t$ becomes large, the system becomes strongly correlated and traditional coupled cluster badly overcorrelates.  Variational coupled cluster theory instead badly undercorrelates.  At half filling, however, UHF becomes rather accurate, energetically, and in fact the UHF energy becomes exact in the $U/t \to \infty$ limit.  Of course SUHF improves upon UHF everywhere, so is also energetically exact for large $U/t$.

We would thus expect the combination of SUHF and coupled cluster to be highly accurate everywhere.  This is borne out by Fig. \ref{Fig:Hubbard}, which shows results for the one-dimensional half-filled Hubbard Hamiltonian with six and ten sites in the left- and right-hand panels, respectively.   Figure \ref{Fig:Hubbard2x4Half} shows results for the two-dimensional Hubbard model where the lattice is $2 \times 4$ and half filled, but with open boundary conditions so as to lift a degeneracy at the Fermi level which is caused by lattice momentum symmetry.  In all three cases, we see that SUVCCSD is roughly equivalent to VCCSD for small $U/t$, while for large $U/t$ it instead resembles SUHF; in the recoupling region it significantly outperforms either.  By ``recoupling region'' we mean values of $U/t$ large enough that the UHF has broken spin symmetry but not so large that the system is effectively described by the N\'eel state in which each lattice site is occupied by a single electron and electrons on adjacent sites have opposite spins.  Adding triples to give SUVCCSDT is exceptionally accurate everywhere for the smaller lattices, but for the 10-site model even higher excitation levels are apparently necessary. 

There is, however, an unpleasant feature here which we should point out.  The errors per electron of both SUHF and VCCSD increase with increasing system size, while UHF gets somewhat better.  In the thermodynamic limit, the errors per electron of UHF and SUHF are the same\cite{PHF} and for larger $U/t$ we would presumably see no improvement of SUVCCSD over UHF.  On the other hand, coupled cluster with singles and doubles based on the UHF reference (UCCSD) improves noticeably over UHF even for fairly large $U/t$ though it does not fully restore either spin symmetry or lattice momentum symmetry (both of which are broken in UHF) while SUVCCSD breaks neither.  Energetically, for very large systems in this particular example there seems to be no reason to go to the expense of using SUVCCSD instead of simply using UCCSD.  The SUVCCSD wave function is presumably more physical in that it has the same symmetries as does the exact ground state wave function, but so too would be a spin-projected UCCSD.  Results for spin projecting the UCCSD wave function will be presented in due time.  Guided by our experience in the Lipkin Hamiltonian,\cite{WahlenStrothman2016} we expect spin projecting UCCSD to be superior in the strongly correlated limit (where it should not be worse than UCCSD itself) but perhaps less accurate in the recoupling region.

\begin{figure*}[t]
\includegraphics[width=0.48\textwidth]{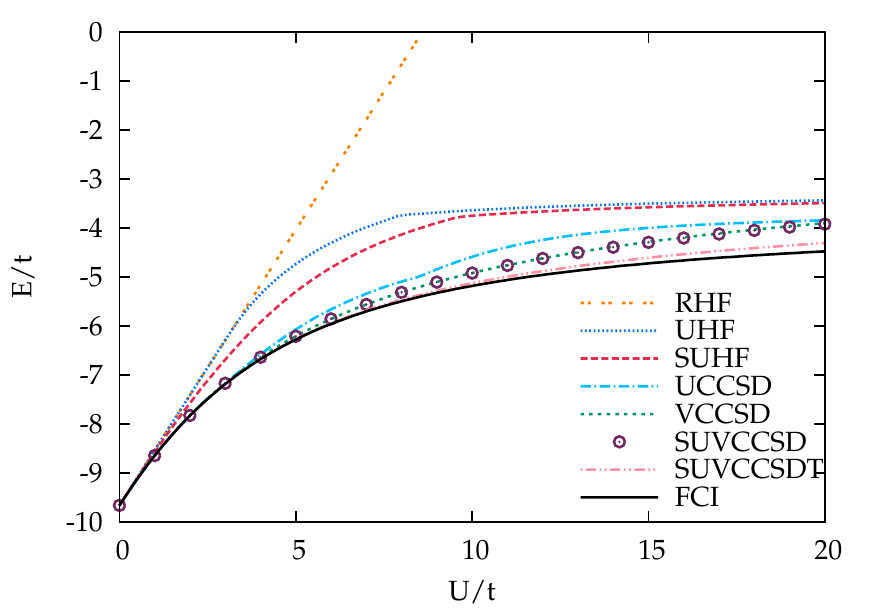}
\hfill
\includegraphics[width=0.48\textwidth]{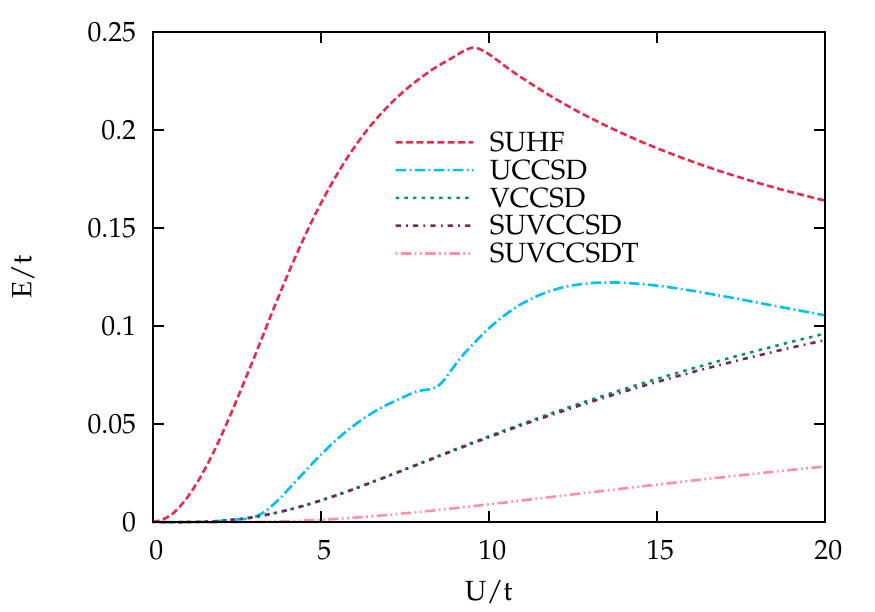}
\caption{Results for the eight-site one-dimensional Hubbard ring with six electrons.  Left panel: Total energies.  Right panel: Errors per electron with respect to the exact result.  Note that the lowest-energy UHF changes character near $U/t \approx 8$ and the SUHF determinant does likewise near $U/t \approx 9.5$.
\label{Fig:Hubbard8}}
\end{figure*}

Away from half filling, UHF and consequently SUHF are less helpful.  In Fig. \ref{Fig:Hubbard8} we show results for an eight-site Hubbard ring with six electrons.  There is a qualitative change in character in the lowest energy UHF as a function of $U/t$, which is reflected in the SUHF, and in fact for large $U/t$ the Hartree-Fock ground state appears to be of generalized Hartree-Fock type (\textit{i.e.} the lowest energy Hartree-Fock breaks both $S^2$ and $S_z$ symmetry).  For this combination of lattice and filling fraction, SUHF is significantly less effective at capturing the strong correlation for large $U/t$, and accordingly VCCSD and SUVCCSD are essentially identical.  The contributions from triple excitations are now very significant for large $U/t$.  Note that there are no triple excitations in SUVCCSD because the PHF polynomial contains only even excitation levels, and single excitations vanish due to momentum symmetry.  Unless these triple excitations can be accounted for, there seems to be little hope of obtaining accurate results.  Results for doped two-dimensional lattices (not shown) display these same basic features.

\begin{figure}[t]
\includegraphics[width=0.75\columnwidth]{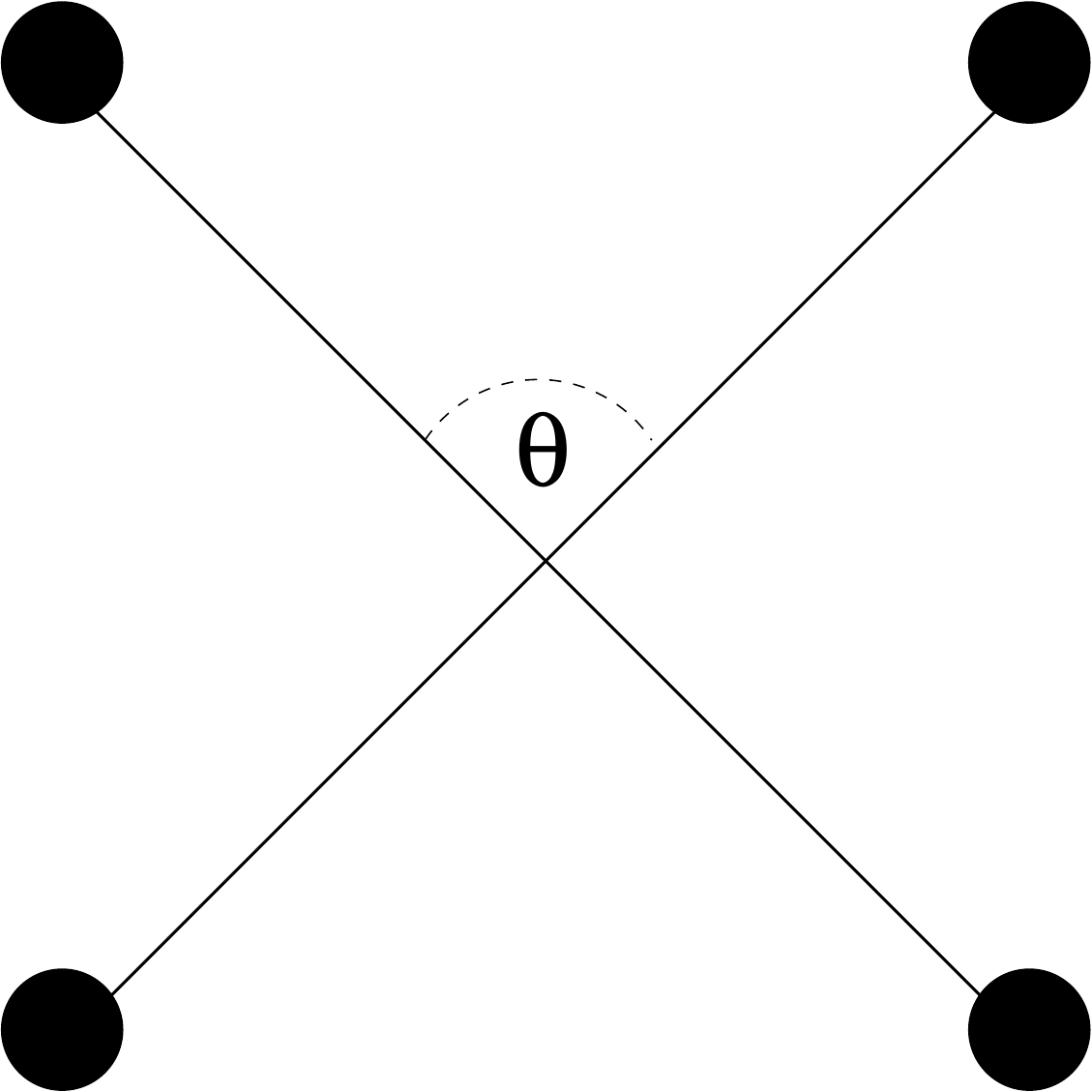}
\caption{Schematic representation of the H$_4$ ring.
\label{Fig:H4Diagram}}
\end{figure}

\begin{figure}[t]
\includegraphics[width=\columnwidth]{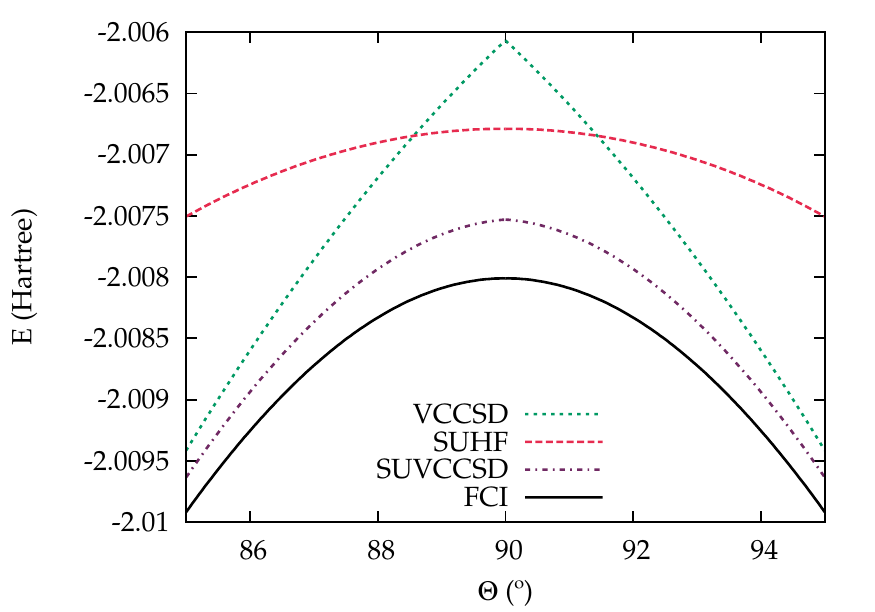}
\caption{Total energies for an H$_4$ ring with radius 3.3 bohr as a function of angle.  We use the cc-pVDZ basis set.  Where VCCSD has a cusp, SUHF and SUVCCSD are correctly smooth.
\label{Fig:H4}}
\end{figure}

\subsection{Molecular Examples}
Let us now turn to a few simple molecular examples.  We begin with the case of four hydrogen atoms on a ring of radius 3.3 bohr, depicted schematically in Fig. \ref{Fig:H4Diagram}.  The ground state and lowest singlet excited state are nearly degenerate for $\theta \approx 90^\circ$, and become exactly degenerate at this high symmetry point.  While the exact curve of the energy as a function of $\theta$ should be smooth with a maximum at $90^\circ$, approximate methods typically instead have a cusp here and some also predict a spurious local minimum.  In  Fig. \ref{Fig:H4} we see that VCCSD is superior to SUHF except near the high symmetry point, where it yields an unphysical cusp.  Combining SUHF and VCCSD in SUVCCSD gives results superior to both methods; there is no cusp, and SUVCCSD is nearly parallel to the exact result.

\begin{figure*}[t]
\includegraphics[width=0.48\textwidth]{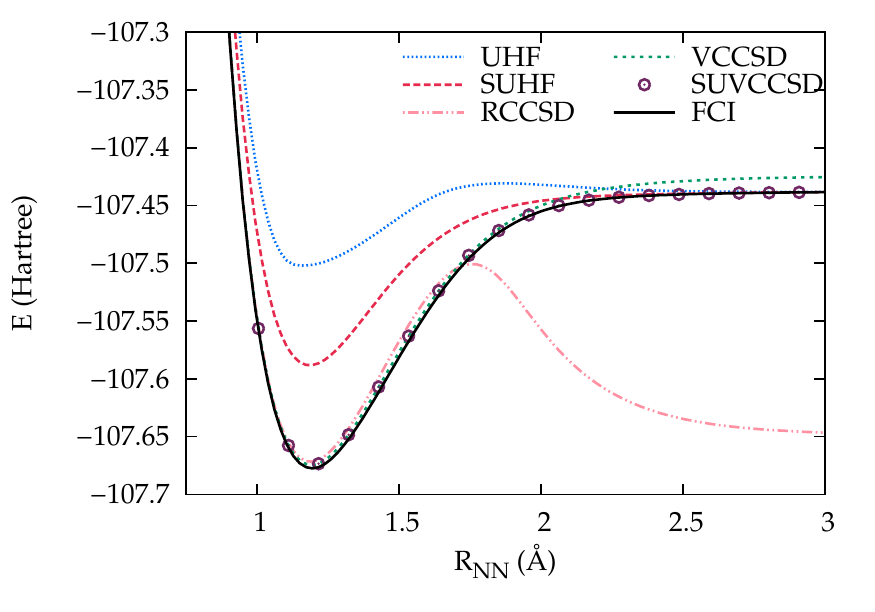}
\hfill
\includegraphics[width=0.48\textwidth]{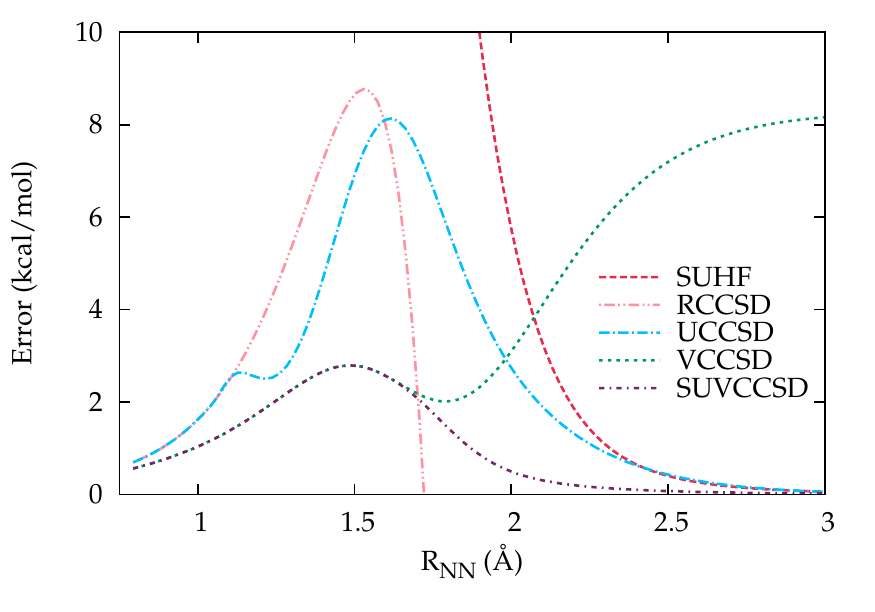}
\caption{Energies in the N$_2$ molecule using the STO-3G basis set.  Left panel: Total energies.  Right panel: Errors with respect to FCI.
\label{Fig:N2}}
\end{figure*}

Next we consider the dissociation of N$_2$ in the STO-3G basis set, as depicted in Fig. \ref{Fig:N2}.  While traditional restricted coupled cluster with single and double excitations (RCCSD) has an unphysical bump and dissociates to an energy which is much too low, VCCSD gives a reasonable curve everywhere.  However, VCCSD does not go to quite the right dissociation limit, while in this minimal basis set, SUHF does.  Accordingly, SUVCCSD follows VCCSD near equilibrium when the system is weakly correlated, yields the right answer at dissociation, and outperforms both VCCSD and SUHF in the intermediate coupling regime.  The right panel of Fig. \ref{Fig:N2} shows errors with respect to full configuration interaction as a function of the N-N bond length $R_\mathrm{NN}$, and we see that in this basis set the maximum error of SUVCCSD is about 2.5 kcal/mol.

\begin{figure*}[t]
\includegraphics[width=0.48\textwidth]{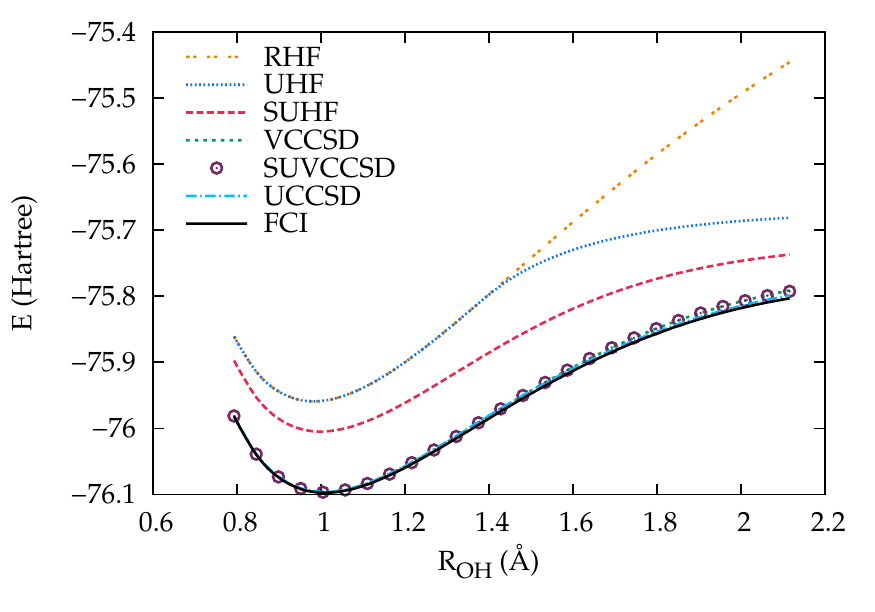}
\hfill
\includegraphics[width=0.48\textwidth]{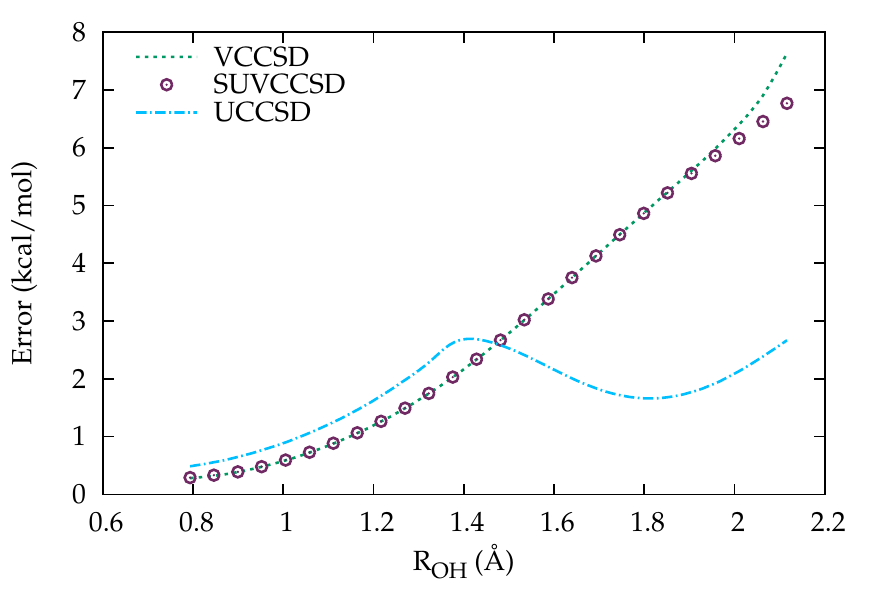}
\caption{Energies in the symmetric double dissociation of H$_2$O using the STO-3G basis set on the hydrogen atoms and the 6-31G basis on the oxygen atom.  Left panel: Total energies.  Right panel: Errors with respect to FCI.
\label{Fig:H2O}}
\end{figure*}

Finally we consider the symmetric double dissociation of H$_2$O, in which we use the STO-3G basis for the hydrogen atoms and the 6-31G basis for the oxygen atom.  We take an H-O-H bond angle of 104.5$^\circ$.  As the left panel of Fig. \ref{Fig:H2O} makes clear, with this slightly larger basis set the SUHF is no longer exact at large bond lengths.  Accordingly, as can be seen from the right panel, SUVCCSD offers only negligible improvement over VCCSD.

\section{Conclusions}
Methods which can accurately describe both strongly and weakly correlated systems remain frustratingly elusive.  Coupled cluster theory is unquestionably successful when correlations are weak, and symmetry projected mean field methods are generally useful when correlations are strong.  Combining the two approaches seems to be a logical approach.  This combination is, however, not as straightforward as one would like because the two theories are formulated in very different ways.  In Ref. \onlinecite{Degroote2016} we showed how to interpolate between coupled cluster and the number-projected Bardeen-Cooper-Schriefer (BCS) state in the pairing or reduced BCS Hamiltonian.  Reference \onlinecite{Qiu2016} formulated spin-projected unrestricted Hartree-Fock in the language of a similarity-transformed Hamiltonion built from particle-hole excitations out of a symmetry-adapted reference.  While we feel that this is an important step toward fruitfully combining PHF and coupled cluster theories, we provided no results for this combination.  Reference \onlinecite{WahlenStrothman2016} shows results for the combination of parity-projected Hartree-Fock and coupled cluster theory when applied to the Lipkin model Hamiltonian.  Here, we provide our first applications to the Hubbard and molecular Hamiltonians, albeit only in a variational formulation.  Thus far, the results seem to be generally encouraging, but much work clearly remains to be done. Most importantly, it seems obvious that the variational approach used here must be abandoned in favor of a similarity-transformation technique.  Work along these lines is underway.  More basically, however, we must address several questions.  Two questions strike us as particularly important.

One important question is under which circumstances combining PHF and coupled cluster will offer any improvements at all.  Our results here are somewhat mixed, but generally we suggest that when PHF is adequate to describe the strong correlations in the system, then combining PHF and coupled cluster should work well.  This should not be too surprising, and is supported by our earlier work.\cite{Degroote2016,WahlenStrothman2016}

We must therefore ask which symmetries should be projectively restored in the first place, and how the symmetry-projected determinant can be expressed in terms of symmetry-adapted particle-hole excitations?  Each symmetry would in general have a different relation between the parameters defining the Thouless transformation that takes us to the broken symmetry determinant on the one hand and the symmetry-adapted particle-hole excitations used to define the similarity-transformed Hamiltonian on the other hand.  Thus, for example, while projecting UHF onto a singlet state gives us the sinh polynomial we have called $F(K_2)$, projecting UHF onto a state with different spin will give us different polynomials.  Projecting spatial symmetry would be different again.

Second, what happens in the thermodynamic limit?  While coupled cluster is extensive (\textit{i.e.} the energy scales properly with system size), extensivity in projected Hartree-Fock is more subtle.  It is true that PHF has an extensive component, in that the energy per particle in the thermodynamic limit is the same as that of the broken symmetry mean field (and therefore below that of the symmetry-adapted mean field from which we begin).  But what happens when we combine PHF and coupled cluster theory?  Certainly we would introduce unlinked terms which in the thermodynamic limit should not be there.

These questions are not easy.  But we are cautiously optimistic that they can be answered and that the resulting methods will have great potential as broadly applicable techniques which can describe all sorts of correlated systems with high accuracy at a computational cost not much different from that of traditional coupled cluster theory.

\section{Supplementary Material}
Supplementary material provides full details for the evaluation of the similarity-transformed energy expression of Eqn. \ref{Eqn:ESPECC} and proves the relation given in Eqn. \ref{Eqn:EpsReln}.

\begin{acknowledgments}
This work was supported by the U.S. Department of Energy, Office of Basic Energy Sciences, Computational and Theoretical Chemistry Program under Award No.DE-FG02-09ER16053. G.E.S. is a Welch Foundation Chair (C-0036).  We would like to acknowledge computational support provided by the Center for the Computational Design of Functional Layered Materials, an Energy Frontier Research Center funded by the U.S. Department of Energy, Office of Science, Basic Energy Sciences under Award DE-SC0012575.
\end{acknowledgments}

\bibliography{PoST}
\end{document}